\documentclass[twocolumn]{article}

\usepackage{afterpage}
\usepackage{amsmath}
\usepackage{amsfonts}
\usepackage{amsthm}
\usepackage{spconf}
\usepackage{floatrow}
\usepackage{graphicx}
\usepackage{subcaption}
\usepackage{placeins}
\newtheorem{proposition}{Proposition}

\graphicspath{{./Figures/}}

\title{Consensus State Gram Matrix Estimation for Stochastic\\ Switching Networks from Spectral Distribution Moments}
\name{Stephen Kruzick and Jos\'{e} M. F. Moura\thanks{
Stephen~Kruzick (skruzick@andrew.cmu.edu) and Jos\'{e}~M.~F.~Moura (moura@andrew.cmu.edu) are with the Department of Electrical and Computer Engineering at Carnegie Mellon University in Pittsburgh, PA, USA.  This work was supported by NSF grant \#~CCF~1513936.}}
\address{Carnegie Mellon University, Department of Electrical Engineering\\5000  Forbes Avenue, Pittsburgh, PA 15213}

\begin{document}
\maketitle

\begin{abstract}
Reaching distributed average consensus quickly and accurately over a network through iterative dynamics represents an important task in numerous distributed applications.  Suitably designed filters applied to the state values can significantly improve the convergence rate.  For constant networks, these filters can be viewed in terms of graph signal processing as polynomials in a single matrix, the consensus iteration matrix, with filter response evaluated at its eigenvalues.  For random, time-varying networks, filter design becomes more complicated, involving eigendecompositions of sums and products of random, time-varying iteration matrices.  This paper focuses on deriving an estimate for the Gram matrix of error in the state vectors over a filtering window for large-scale, stationary, switching random networks. The result depends on the moments of the empirical spectral distribution, which can be estimated through Monte-Carlo simulation.  This work then defines a quadratic objective function to minimize the expected consensus estimate error norm.  Simulation results provide support for the approximation.
\end{abstract}


\section{Introduction}\label{Introduction}
In networks, reaching agreement on the mean of data spread among the nodes without the mediation of a leader node and only using local communications represents a common, extensively studied task.  This problem, known as distributed average consensus~\cite{OSab2}, has applications such as sensor data fusion~\cite{LXia2}, processor load balancing~\cite{GCyb1}, flocking of multiagent systems~\cite{OSab1}, and distributed inference~\cite{SKar2}.  An iterative algorithm to achieve this goal can be implemented as a discrete dynamic system, where each node maintains a state initially set to its data value and updated as a linear function of neighboring states at each iteration.  Collecting all states at iteration $n$ into vector $\mathbf{x}_n$ with initial data $\mathbf{x}_0$, this can be described by the iteration equation
\begin{equation}
\mathbf{x}_{n+1}=W\left(\mathcal{G}\right)\mathbf{x}_n 
\end{equation}
in which $W\left(\mathcal{G}\right)$ is an iteration matrix that respects the local structure of the network graph $\mathcal{G}$.  Provided $W\left(\mathcal{G}\right)$ satisfies
\begin{equation}
{\boldsymbol\ell}^\top W\left(\mathcal{G}\right)={\boldsymbol\ell}^\top,\enskip W\left(\mathcal{G}\right)\mathbf{1}=\mathbf{1}, \enskip \rho\left(W\left(\mathcal{G}\right)-J_{\boldsymbol\ell}\right)<1
\end{equation}
where $\mathbf{1}$ is the vector of ones, $J_{\boldsymbol\ell}=\mathbf{1}{\boldsymbol\ell}^\top/{\boldsymbol\ell}^\top\mathbf{1}$, and $\rho$ is the spectral radius, the state $\mathbf{x}_n$ asymptotically converges at exponential rate to $\lim_{n\rightarrow\infty}\mathbf{x}_n=J_{\boldsymbol\ell}\mathbf{x}_0$~\cite{SKar2}. The result is the unweighted average consensus vector when ${\boldsymbol\ell}=\mathbf{1}$, and the exponential rate of convergence relates to $\ln\left(\rho\left(W\left(\mathcal{G}\right)-J_{\mathbf{\ell}}\right)\right)$~\cite{SKar2}.

To achieve more accurate results or to require fewer communication iterations for given accuracy, rapid convergence is preferred.  Faster consensus systems have been approached by designing the iteration matrix $W$ given the network topology $\mathcal{G}$~\cite{LXia1}, designing the network topology $\mathcal{G}$ given a weight matrix scheme $W\left(\mathcal{G}\right)$~\cite{SKar1}, and by modifying the algorithm through application of filters incorporating previous states.  For known network topology, consensus can be exactly reached in finitely many iterations through application of a filter related to the minimal polynomial of the iteration matrix~\cite{SSun1,ASan4}, requiring degree $K-1$ where $K$ is the number of distinct iteration matrix eigenvalues.  To achieve fast asymptotic consensus, periodically applied filters of lower degree were designed by means of a semidefinite program in~\cite{EKok1}.  Other approaches to filter design for accelerated consensus include those in~\cite{EMon1,ALou1,SApe1,FGam1}.

In terms of graph signal processing, consensus acceleration filters can be interpreted (for constant network topology) as lowpass graph filters that attempt to eliminate signal content in all eigenspaces except corresponding to the consensus eigenvector $\mathbf{v}=\mathbf{1}$ with eigenvalue $\lambda(W)=1$.  Hence, knowledge of the iteration matrix eigenvalues, when available, can be applied to design these filters.  For random networks with random iteration matrices, useful information can sometimes be obtained for large-scale problems through asymptotic theorems from random matrix theory.  The empirical spectral distribution and empirical spectral density of matrix $W$ are respectively defined as 
\begin{align}
F_W\left(x\right)&=\textstyle\frac{1}{N}\sum_{k=1}^{k=N}\chi\left(x-\lambda_{k}\left(W\right)\right) \\
f_W\left(x\right)&=\textstyle\frac{1}{N}\sum_{k=1}^{k=N}\delta\left(x-\lambda_{k}\left(W\right)\right)
\end{align}
where $\chi$ is an indicator function and $\delta$ is the Dirac delta function~\cite{RCou1}.  While these are random distribution and density functions for a random matrix $W$, sometimes they may have limiting behavior as the matrix size increases~\cite{RCou1,VGir1,SVer1}.  Deterministic approximations for the empirical spectral density were used to define intervals for consensus acceleration filter response optimization in~\cite{SKru1,SKru2,SKru3,SKru4}, achieving performance gains for certain constant, large-scale network models through Chebyshev filter design.  However, the graph signal processing intuition breaks down when the network and corresponding iteration matrices become time-varying.  Intuitively, the deterministic information regarding the empirical distribution should remain useful for filter design when the network stochastic process varies slowly and has the same marginal distribution at each time instant, becoming less relevant as the network changes faster.  

This paper proposes design criteria for consensus acceleration filters on random switching networks, a relatively simple class of time-varying network models, for certain large-scale random network distributions.  Each network is drawn from some specified random network distribution.  At each time iteration, the network either remains constant or switches to a new, independent sample from the random network distribution according to a Bernoulli trial with fixed probability.  The proposed quadratic optimization objective involves an approximation of the expected Gram matrix of error in the state vectors over a filtering window.  The derived approximation depends only on the moments of the expected empirical spectral distribution of the iteration matrices and on the switching probability.  Section~\ref{Methods} derives the approximation and presents the proposed optimization problem.  Section~\ref{Simulations} supports the proposed methods with simulations showing the approximation quality.  Finally, Section~\ref{Conclusion} provides concluding analysis.

\section{Gram Matrix Approximation and\\ Filter Design Method}\label{Methods}
Consider distributed average consensus with respect to a time-varying sequence of iteration matrices $\left\{W_n\right\}$ arising from a random switching network with $N$ nodes and switching probability $p_{\textrm{sw}}$.  To compute the mean of the initial data $\mathbf{x}_0$, the network implements a dynamic system with state $\mathbf{x}_n$ at time iteration $n$ described by
\begin{equation}
\mathbf{x}_n=W_n\mathbf{x}_{n-1}.
\end{equation}
Provided the iteration matrices $W_n$ satisfy the consensus conditions
\begin{equation}
\mathbf{1}^\top W_n=\mathbf{1}^\top,\quad W_n\mathbf{1}=\mathbf{1}, \quad \rho\left(W_n-J_{\mathbf{1}}\right)<1,\label{ConsCond}
\end{equation}
where $J_{\mathbf{1}}=\mathbf{1}\mathbf{1}^\top/N$, the state vector converges to a constant equal to the average $\mathbf{\overline{x}}_0$ of the initial values.  That is, 
\begin{equation}
\lim_{n\rightarrow\infty}\mathbf{x}_n=J\mathbf{x_0}=\left(\mathbf{\overline{x}}_0\right)\mathbf{1}.
\end{equation}
This paper examines undirected graphs and employs the iteration matrix scheme $W_{n}= I-\alpha L\left(\mathcal{G}_{n}\right)$, which satisfies the properties in~\eqref{ConsCond} when the network graphs $\left\{\mathcal{G}_{n}\right\}$ are connected and $\alpha$ is chosen suitably.  To improve the convergence rate, a degree $d$ filter with coefficients $\left\{\mathbf{a}_{k+1}\right\}_{k=0}^{k=d}$ will be periodically applied to previous state values to update the current state vector according to the following equation.
\begin{equation}
\mathbf{x}_n:=\sum_{k=0}^{k=d}\mathbf{a}_{k+1}\mathbf{x}_{n-d+k}, \quad n\equiv 0 \left(\textrm{mod }d\right)
\end{equation}
Thus, for initial vector $\mathbf{x}_{0}$ the state vector terms used for filtering are given by $\mathbf{x}_k=\phi_k\left(\left\{W_n\right\}_{n=1}^{n=d}\right)\mathbf{x}_0$ where
\begin{equation}
\phi_k\left(\left\{W_n\right\}_{n=1}^{n=d}\right)=W_{k}\cdots W_{1}, \enskip
\phi_0\left(\left\{W_n\right\}_{n=1}^{n=d}\right)=I
\end{equation}
for $k=0,\ldots,d$. 
Because each iteration matrix has eigenvalue $\lambda=1$ corresponding to the consensus eigenvector $\mathbf{1}$, the filter coefficients must have unit sum to preserve the signal mean. Collecting the filter coefficients into a vector $\mathbf{a}$, this constraint can be expressed as $\mathbf{1}^\top\mathbf{a}=1$.

Attempting to directly optimize the expected norm of the filter output error for the worst case input proves challenging.  Instead, the filters designed in this section approximately minimize the expected norm of the filter output error vector with respect to the random iteration matrix sequence and with respect to the initial error vector.  Let $\mathbf{x}_0=\overline{\mathbf{x}}_0\mathbf{1}+\mathbf{v}$ where $\mathbf{v}$ is orthogonal to $\mathbf{1}$, and assume for simplicity that $\mathbf{v}$ is uniformly distributed on unit norm vectors orthogonal to $\mathbf{1}$.  By Jensen's inequality, the square root of the expected norm squared provides a lower bound for the expected norm.  Thus, rather than minimizing the expected norm of the filter error directly, the expected norm squared will be minimized as described in~\eqref{Opt1}, where $\mathbf{v}$ is uniformly distributed on $\left\{\mathbf{v}\in\mathbb{R}^N | \mathbf{v}\bot \mathbf{1},\left\|\mathbf{v}\right\|=1\right\}$.
\begin{equation}
\begin{aligned}
\min_{\mathbf{a}} \quad\!& \operatorname{E}_{\left\{W_n\right\},\mathbf{v}}\!\!\left[\!\left\|\sum_{k=0}^{k=d}\mathbf{a}_{k+1}\phi_{k}\left(\left\{W_{n}\right\}_{n=1}^{n=d}\right)\mathbf{v}\right\|_{\!2}^{\!2}\right]\\
\textrm{s.t.} \quad\!& \mathbf{1}^\top\mathbf{a}=1
\end{aligned}\label{Opt1}
\end{equation}

Denote by $\mathbf{s}$ the network switching sequence where $\mathbf{s}_{1}=1$ and, for all $n>1$, $\mathbf{s}_{n}=1$ if $W_{n}=W_{n-1}$ and $\mathbf{s}_{n}=0$ otherwise.  That is, $\mathbf{s}_n$ determines whether $W_n$ is a new iteration matrix (within the filtering window), with $W_1$ always counted.  The space of all possible switching sequences are the $d$-tuples 
\begin{equation}
S_d=\left\{1\right\}\times\left\{0,1\right\}\times\cdots\times\left\{0,1\right\},
\end{equation}
and the total number of independent networks is
\begin{equation}
\#\mathbf{s}=\textstyle\sum_{k=1}^{k=d}\mathbf{s}_k, \qquad \mathbf{s}\in S_{d}
\end{equation}
with $\#\mathbf{s}-1$ total switching events.  For a switching process governed by independent Bernoulli trials with switching probability $p_{sw}$, switching sequence $\mathbf{s}\in S_d$ has probability mass given by
\begin{equation}
p\left(\mathbf{s}\right)=\left(p_{\textrm{sw}}\right)^{\left(\#\mathbf{s}-1\right)}\left(1-p_{\textrm{sw}}\right)^{\left(d-1\right)-\left(\#\mathbf{s}-1\right)}.
\end{equation}

The distribution of the iteration matrix sequence $\left\{W_{n}\right\}_{n=1}^{n=d}$ can be factored into the distribution of the switching sequence $\mathbf{s}$ and distribution of $\left\{W_{n}\left(\mathbf{s}\right)\right\}_{n=1}^{n=d}$ given the switching sequence.  Writing the expected norm squared as an inner product, the optimization problem in~\eqref{Opt1} becomes
\begin{equation}
\begin{aligned}
\min_{\mathbf{a}} \quad& \mathbf{a}^\top 
\left(\sum_{\mathbf{s}\in S_d} p\left(\mathbf{s}\right) \operatorname{E}_{\left\{W_n\left(\mathbf{s}\right)\right\},\mathbf{v}}\left[
{Q}\left(\mathbf{s}\right)|\mathbf{s}\right]\right)\mathbf{a}\\
\textrm{s.t.} \quad& \mathbf{1}^\top\mathbf{a}=1
\end{aligned}\label{Opt2}
\end{equation}
where $Q\left(\mathbf{s}\right)$ is a $\left(d+1\right)\times \left(d+1\right)$ random Gram matrix defined for each sequence $\mathbf{s}\in S_d$ by entries
\begin{equation}
\begin{aligned}
\!\!Q_{ij}\left(\mathbf{s}\right)\!=&\left\langle 
\phi_{i-1}\left(\left\{W_{n}\left(\mathbf{s}\right)\right\}_{n=1}^{n=d}\right)\mathbf{v} , 
\right. \\ &\hspace{4pt}\left.
\phi_{j-1}\left(\left\{W_{n}\left(\mathbf{s}\right)\right\}_{n=1}^{n=d}\right)\mathbf{v} \right\rangle\!.
\end{aligned}
\end{equation}

The above optimization problem could be used for filter design by computing the values of $\operatorname{E}_{\left\{W_n\left(\mathbf{s}\right)\right\},\mathbf{v}}\left[
{Q}\left(\mathbf{s}\right)|\mathbf{s}\right]$ through simulation.  However, the intent of this paper is to connect information regarding the empirical spectral distribution of the iteration matrices to filter design for large-scale random switching networks, as done for constant random networks in~\cite{SKru3,SKru4}.  Therefore, an analytical estimate (under suitable conditions) based on a deterministic approximation of the expected empirical spectral distribution will be described below.

Before proceeding, some notation must first be introduced.  Note that each switching sequence $\mathbf{s}$ corresponds to an integer composition of $d$ with $\#\mathbf{s}$ partitions, namely $c\left(\mathbf{s}\right)=\left(c_1\left(\mathbf{s}\right),\ldots,c_{\#\mathbf{s}}\left(\mathbf{s}\right)\right)$  where each $c_m\left(\mathbf{s}\right)$ is the number of iterations the $m$th network is used before switching.  Let $c'_m\left(\mathbf{s},n\right)$ be the number of iterations the $m$th network is used up to (and including) iteration $n\geq1$ with $c'_m\left(\mathbf{s},0\right)=0$.  More explicitly,
\begin{equation}
	c'_m\left(\mathbf{s},n\right)=\left\{\!\!\!\!\begin{array}{cc}0 & n<\sum_{k=1}^{m-1} c_k\left(\mathbf{s}\right) \\ c_m\left(\mathbf{s}\right) & n>\sum_{k=1}^{m} c_k\left(\mathbf{s}\right)  \\ n-\sum_{k=1}^{m-1} c_k\left(\mathbf{s}\right) & \textrm{otherwise} \end{array}\right. \!\!\!\!.
\end{equation}

Let $\left\{\mathbf{u}_{n,k}\left(\mathbf{s}\right)\right\}_{k=1}^{k=N-1}\cup\left\{\mathbf{1}/\sqrt{N}\right\}$ be the orthonormal basis of eigenvectors for $W_n\left(\mathbf{s}\right)$ with $\left\{\mathbf{u}_{0,k}\left(\mathbf{s}\right)\right\}_{k=1}^{k=N}$ the standard basis for $\mathbb{R}^N$.  Note that if the network does not switch, the basis of eigenvectors remains the same.  Assume that the eigenvectors before and after a random network switch have the following properties.  
\begin{gather}
\operatorname{E}\left[\left\langle \mathbf{v},\mathbf{u}_{0,k}(\mathbf{s})\right\rangle^2\right]=1/N\\
\operatorname{E}\left[\left\langle \mathbf{u}_{m,i}(\mathbf{s}),\mathbf{u}_{n,j}(\mathbf{s})\right\rangle^2\right]=1/(N-1)\\
\operatorname{E}\left[\left\langle \mathbf{u}_{m,i}(\mathbf{s}),\mathbf{u}_{n,j}(\mathbf{s})\right\rangle\left\langle \mathbf{u}_{m,i}(\mathbf{s}),\mathbf{u}_{n,k}(\mathbf{s})\right\rangle\right]=0
\end{gather}
Then $\operatorname{E}_{\left\{W_n\left(\mathbf{s}\right)\right\},\mathbf{v}}\left[
{Q}\left(\mathbf{s}\right)|\mathbf{s}\right]$ can be approximated by the following expression, where $\widehat{f}_W$ is the expected empirical spectral density of the iteration matrix.  
\begin{equation}
\widehat{Q}_{ij}\left(\mathbf{s}\right)=\prod_{m=\hphantom{\#}1}^{m=\#\mathbf{s}} \operatorname{E}_{\widehat{f}_W}\left[\lambda^{c'_m\left(\mathbf{s},i-1\right)+c'_m\left(\mathbf{s},j-1\right)}\right]\label{EstQ}
\end{equation}
The expected value of \eqref{EstQ} with respect to the switching sequence then provides the approximation of $\operatorname{E}\left[Q\right]$, where $Q$ is the random Gram matrix of error in the state vectors (not given $\mathbf{s}$).
\begin{equation}
\widehat{Q}=\operatorname{E}_\mathbf{s}\left[\widehat{Q}\left(\mathbf{s}\right)\right]=\sum_{\mathbf{s}\in S_d} p\left(\mathbf{s}\right)\widehat{Q}\left(\mathbf{s}\right)\label{EstQ2}
\end{equation}

A brief outline for the derivation of~\eqref{EstQ} will now be presented.  The above result follows from expressing state terms $x_i,x_j$ in the $d$th eigenbasis through a sequence of projections, writing the inner product, and applying the stated eigenbasis assumptions.  Those steps yield a result that depends only on the moments of the true empirical spectral density of each independent iteration matrix.  As an approximation, the moments of the expected empirical spectral density are substituted for the true moments.  Full derivation details will be published in an extended version of this paper.

\begin{proposition}
The estimate matrix $\widehat{Q}\left(\mathbf{s}\right)$ defined in~\eqref{EstQ} is positive semidefinite for any density $\widehat{f}_W$ and switching sequence $\mathbf{s}\in S_d$.  Consequently, the optimization problem in~\eqref{PropLCQP} is a positive definite linearly constrained quadratic program (LCQP).
\end{proposition}

While full details are omitted, proof of the above proposition follows from representing $\widehat{Q}\left(\mathbf{s}\right)$ as an expected Gram matrix.  (Note that this is non-trivial, since $\widehat{Q}(\mathbf{s})$ is not exactly $\operatorname{E}\left[Q(\mathbf{s})|\mathbf{s}\right]$.)  Substitution of the computed value of $\widehat{Q}\left(\mathbf{s}\right)$ for $\operatorname{E}_{\left\{W_n\left(\mathbf{s}\right)\right\},\mathbf{v}}\left[
{Q}\left(\mathbf{s}\right)|\mathbf{s}\right]$ in equation~\eqref{Opt2} results in the final form of the optimization problem.
\begin{equation}
\begin{aligned}
\min_{\mathbf{a}} \quad& \mathbf{a}^\top 
\left(\sum_{\mathbf{s}\in S_{d}} p\left(\mathbf{s}\right) 
\widehat{Q}\left(\mathbf{s}\right)\right)\mathbf{a}\\
\textrm{s.t.} \quad& \mathbf{1}^\top\mathbf{a}=1
\end{aligned}\label{PropLCQP}
\end{equation}
This formulation is a LCQP in which all matrices are positive semidefinite.  
Section~\ref{Simulations} provides simulations demonstrating that a good approximation of the expected gram matrix is achieved for several network distributions, and results for filters designed according to the above optimization problem will appear in an extended version of this paper.



\begin{figure*}
\ffigbox[\textwidth]{
\begin{subfloatrow}[3]

\ffigbox[.305\textwidth]{
\includegraphics[width=\linewidth]{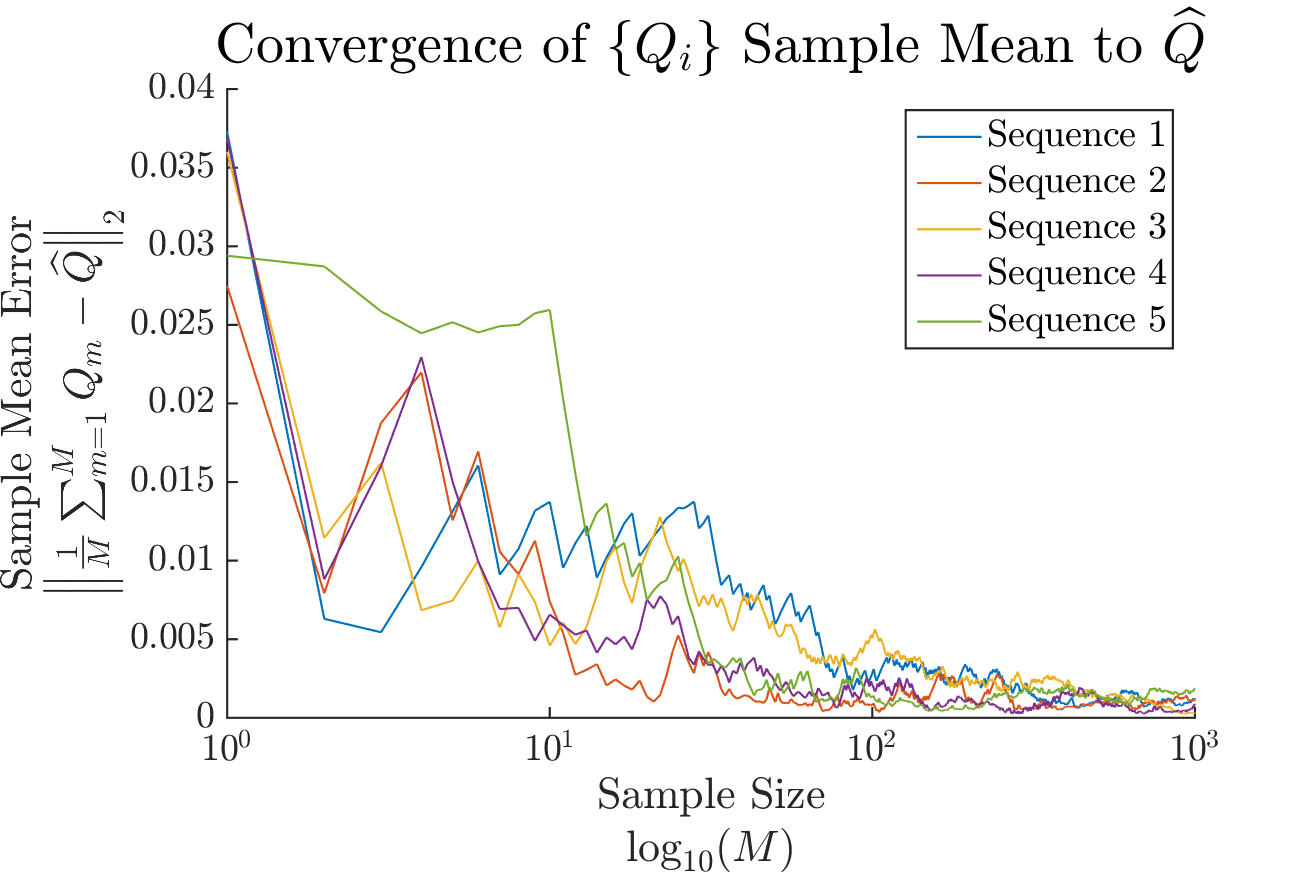}}{\caption{Erd\H{o}s-R\'{e}nyi network ($N=1000$ nodes, link prob. $p=0.03$, switch prob. $p_{\textrm{sw}}=0.4$, filter deg. $d=5$)}\label{SimApprox1}}

\ffigbox[.305\textwidth]{
\includegraphics[width=\linewidth]{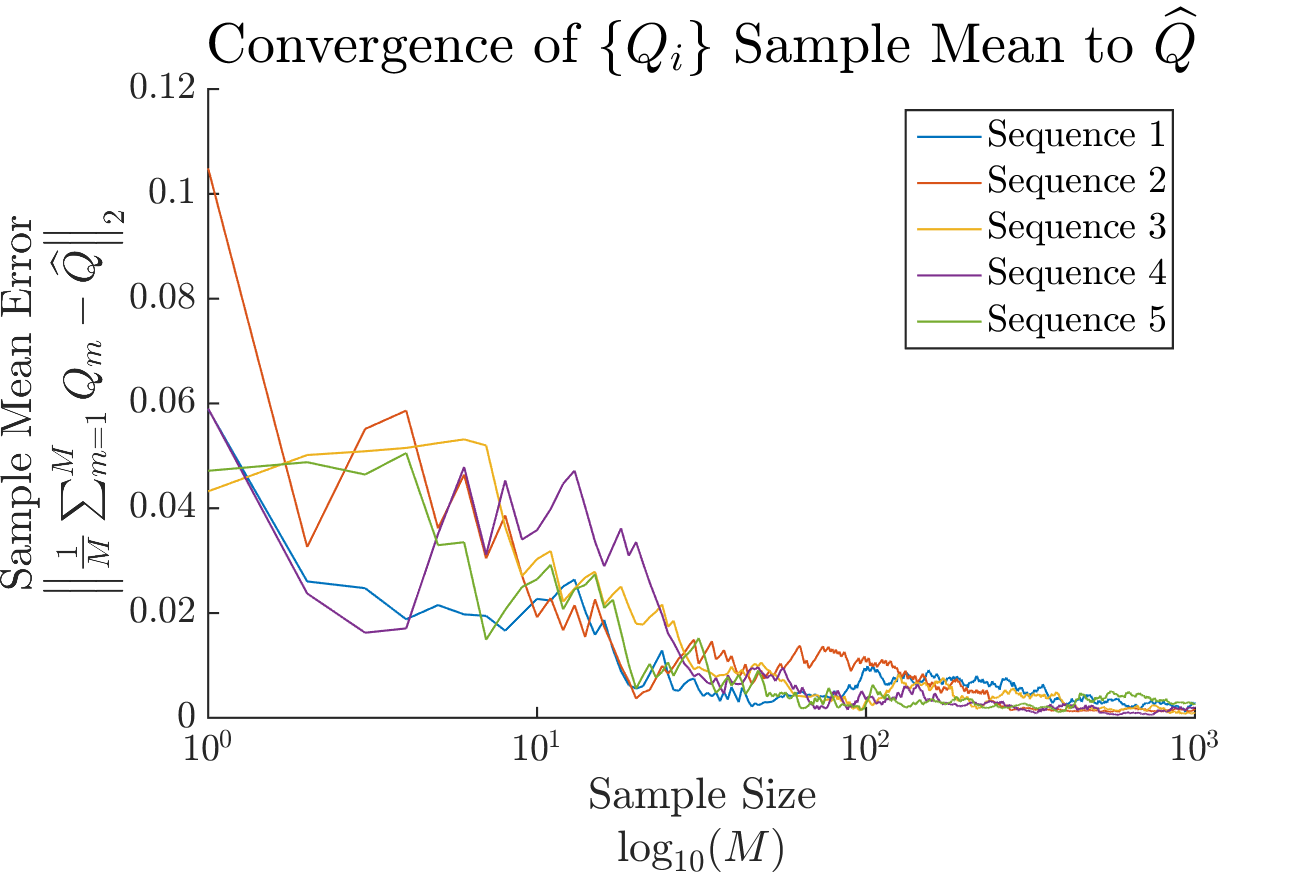}}{\caption{Random location network ($N=1000$ nodes, link rad. $r=1.2\sqrt{\log \left(N\right)/N}$, switch prob. $p_{\textrm{sw}}=0.6$, filter deg. $d=3$)}\label{SimApprox2}}

\ffigbox[.305\textwidth]{
\includegraphics[width=\linewidth]{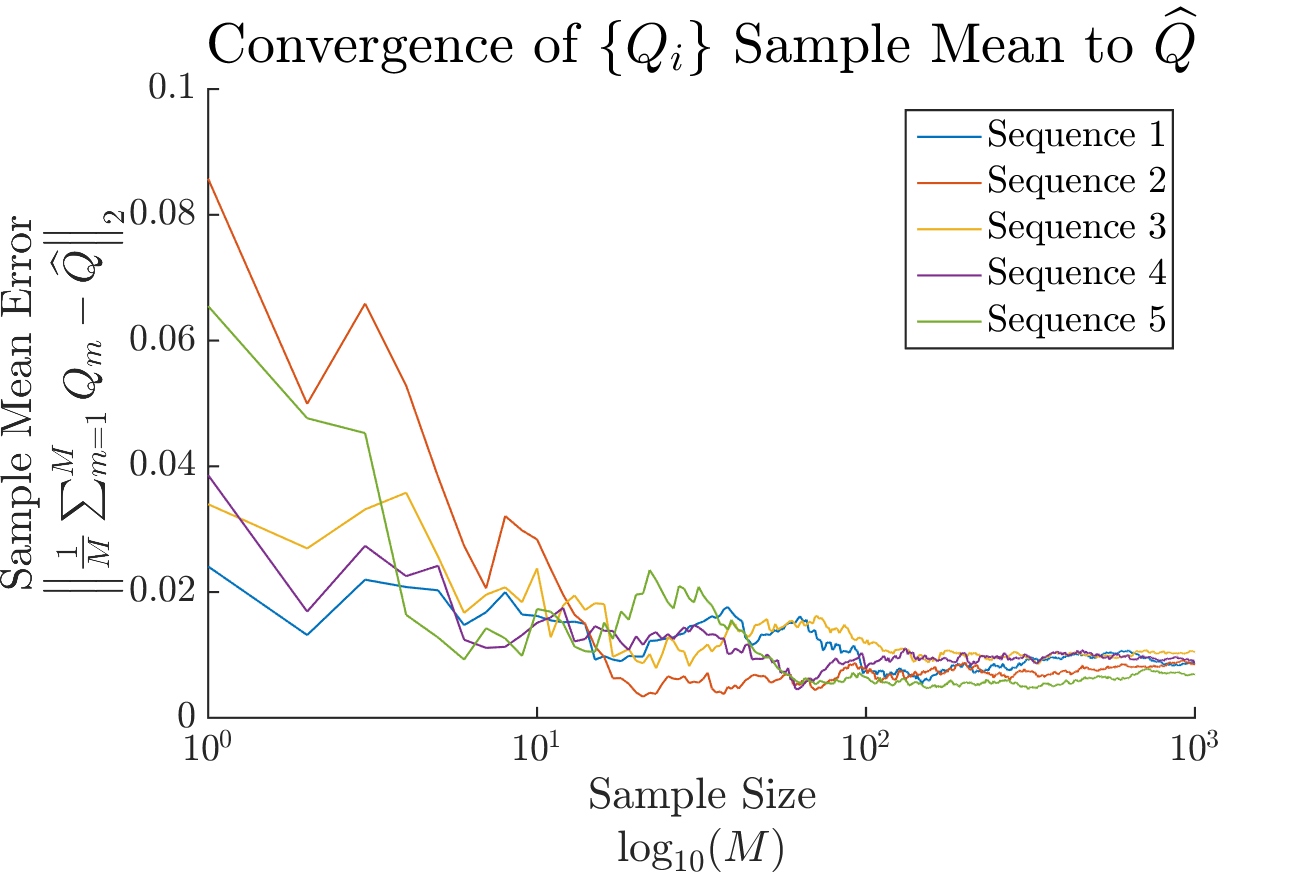}}{\caption{SBM network ($N=600$ nodes, populations $(100,200,300)$, link probs. $p=0.04$ (same pop.) $p=0.02$ (different pop.), switch prob. $p=0.1$, filter deg. $d=4$) \vspace{-2pt}}\label{SimApprox3}}

\end{subfloatrow}}
{\caption{Estimation error (spectral norm) between the approximate Gram matrix $\widehat{Q}$ and the sample mean of independently generated true Gram matrices $\left\{Q_m\right\}_{m=1}^{m=M}$ for increasing sample size $M$ plotted for (a) Erd\H{o}s-R\'{e}nyi network, (b) random location network, and (c) stochastic block model network (model parameters, switching probabilities, and filter degrees listed).  Each figure shows plots for five independent sample sets.\vspace{-10pt}} 
\label{SimFig1}}

\end{figure*}

\section{Simulations}\label{Simulations}
To demonstrate the efficacy of the proposed Gram matrix approximation, this section provides simulations for three undirected random network models with random switching.  Specifically, results shown examine Erd\H{o}s-R\'{e}nyi~\cite{EGil1}, stochastic block model~\cite{KAvr1}, and random location~\cite{PGup1} networks that switch to a new independent random network based on an independent Bernoulli trial at each iteration, with fixed switching probability.  For Erd\H{o}s-R\'{e}nyi networks with $N$ nodes, each pair of nodes forms a link independently with identical fixed probability~\cite{EGil1}.  For stochastic block models with $N$ nodes partitioned into $M$ populations with sizes $(N_1,\ldots, N_M)$, each pair of nodes forms a link with probability depending on the two containing populations~\cite{KAvr1}.  For random location networks with $N$ nodes, the nodes are independently placed uniformly in a unit area and form links if within a given communication radius that produces connectedness with high probability~\cite{PGup1}.

The simulations results in Figure~\ref{SimFig1} demonstrate that $\widehat{Q}$ approximates the expected Gram matrix of error in the state vectors. Each plot shows the spectral norm of the difference between $\widehat{Q}=\operatorname{E}_{\mathbf{s}}\left[\widehat{Q}\left(\mathbf{s}\right)|\mathbf{s}\right]$ from~\eqref{EstQ}-\eqref{EstQ2} and the sample mean of independently generated Gram matrices $\left\{Q_m\right\}_{m=1}^{m=M}$ as the sample size $M$ increases.  Specifically, Figures~\ref{SimApprox1}-\ref{SimApprox3} each plot the error norm for five independent sample sets with maximum size $1000$ for the network models described above with various parameters and filter degrees.  The moments of the  expected empirical spectral density of the weight matrix for each random network model (marginally at any time instant) were estimated via Monte-Carlo simulation over $1000$ trials, allowing application of~\eqref{EstQ} to compute $\widehat{Q}$.  A sample of $1000$ Gram matrices for filter terms were independently generated by drawing initial data $\mathbf{x}_0$ uniformly from $\left\{\mathbf{v}\in\mathbb{R}^N|\mathbf{v}\bot\mathbf{1}, \left\|\mathbf{v}\right\|_2=1\right\}$ and independently generating sequences of weight matrices $\left\{W_n\right\}_{n=1}^{n=d}$ corresponding to graphs drawn from the random network distribution according to $W_n=I-\alpha L\left(G_n\right)$ (where $\alpha$ is approximately optimal for the network model) to produce filter terms $\mathbf{x}_0,\ldots, \mathbf{x}_d$.  The results indicate good approximation quality for the Erd\H{o}s-R\'{e}nyi model in Figure~\ref{SimApprox1} and random location network model in Figure~\ref{SimApprox2}.  In these cases, the assumptions from Section~\ref{Methods} regarding the eigenvectors are suitable as a simplifying approximation due to the high symmetry of the network model with respect to node permutations.  The approximation achieved for the stochastic block model in Figure~\ref{SimApprox3} is of lower quality, as the assumptions from Section~\ref{Methods} regarding the eigenvectors are clearly violated.  Nevertheless, the approximation may remain suitable for filter design.


\section{Conclusion}\label{Conclusion}
This paper derived an approximation to the expected Gram matrix of the error in state vector terms over a time interval for a consensus system on switching random networks.  Under certain assumptions regarding the random eigenvectors, the resulting positive semidefinite matrix depends only on the switching probability and on the moments of the expected empirical spectral distribution.  This approximate expected Gram matrix was used to define a quadratic program that designs filters minimizing the norm of the consensus estimate error in expectation with respect to the network and with respect to the input vector.  Simulation results provided for Erd\H{o}s-R\'{e}nyi, random location, and stochastic block model random networks with stochastic switching demonstrate good approximation quality.  
Continuing work will focus on directly minimizing the spectral radius.  Additionally, future efforts will extend analysis to more involved random network models, such as networks with directed links and time varying networks with statistical dependency as the network changes.

\bibliographystyle{IEEEtran}
\bibliography{Asilomar2017_References}

\end{document}